\documentclass[aps,pra,preprint,showpacs]{revtex4-1}

\usepackage{hyperref, soul}
\usepackage{graphicx}
\usepackage{amsfonts,amsmath,amssymb}
\usepackage{appendix}
\usepackage[latin1]{inputenc}
\usepackage{color}
\usepackage{ulem}

\graphicspath{fig/}

\definecolor{forestgreen}{RGB}{34, 139, 34}

\begin{document}

\title{Isotope shifts of natural Sr$^+$ measured by laser fluorescence in a sympathetically cooled Coulomb crystal}

\author{B. Dubost}
\altaffiliation{also at ICFO-Institut de Ciencies Fotoniques, Mediterranean Technology Park, 08860 Castelldefels (Barcelona), Spain.}

\author{R. Dubessy}
\altaffiliation{Permanent address: Laboratoire de Physique des Lasers, \\CNRS-UMR7538, Universit{\'e} Paris 13--Institut Galil{\'e}e, Villetaneuse, France}

\author{B. Szymanski}

\author{S. Guibal}

\author{J.-P. Likforman}

\author{L. Guidoni}
\email{luca.guidoni@univ-paris-diderot.fr}
\affiliation{Université Paris--Diderot, Sorbonne Paris Cité, Laboratoire Matériaux et Phénomènes Quantiques,\\ UMR 7162 CNRS, F-75205 Paris, France}


\begin{abstract} 
We measured by laser spectroscopy the isotope shifts between naturally-occurring even-isotopes of strontium ions for both the  $5s\,\,^2S_{1/2}\to 5p\,\,^2P_{1/2}$  (violet) and the $4d\,\,^2D_{3/2}\to 5p\,\,^2P_{1/2}$ (infrared) dipole-allowed optical transitions.
Fluorescence spectra were taken by simultaneous measurements on a two-component Coulomb crystal in a linear Paul trap containing $10^3$--$10^4$ laser-cooled Sr$^+$ ions.
The isotope shifts are extracted from the experimental spectra by fitting the data with the analytical solution of the optical Bloch equations describing a three-level atom in interaction with two laser beams.
This technique allowed us to increase the precision with respect to previously reported data obtained by optogalvanic spectroscopy or fast atomic-beam techniques.
The results for the $5s\,\,^2S_{1/2}\to 5p\,\,^2P_{1/2}$ transition are $\nu_{88}-\nu_{84}=+378(4)$~MHz and $\nu_{88}-\nu_{86}=+170(3)$~MHz, in agreement with previously reported measurements.
In the case of the previously unexplored $4d\,\,^2D_{3/2}\to 5p\,\,^2P_{1/2}$ transition we find $\nu_{88}-\nu_{84}=-828(4)$~MHz and $\nu_{88}-\nu_{86}=-402(2)$~MHz.
These results provide more data for stringent tests of theoretical calculations of the isotope shifts of alkali-metal-like atoms.
Moreover, they simplify the identification and the addressing of Sr$^+$ isotopes for ion frequency standards or quantum-information-processing applications in the case of multi-isotope ion strings.

\end{abstract}

\maketitle

\section{Introduction}
\label{sec:intro}
Isotope shift measurements by optical spectroscopy provide information about nuclear structure and constitute an important complement to nuclear-physics experiments that investigate nuclear-charge distribution (e.g. muonic x-ray isotope shifts or electron-scattering) \cite{King:1984}.
The case of strontium ($Z=38$) is particularly studied because it belongs to the elements close to the $Z=40$ subshell closure that causes rapid variation of the nuclear properties as a function of the neutron number \cite{Buchinger:1990}.
In order to extract information about the nucleus, spectroscopic data (i.e. isotope shifts and hyperfine-splitting) have to be compared to some theoretical model, able to evaluate the ``electronic factors'' that take into account the effect of nuclear charge distributions on the electronic wave-functions.
For the alkali-earth elements, these factors are calculated in an easier way for the singly-ionized state, with a single electron in the outer shell. 
During the last years, impressive progress has been made in performing these calculations that estimate hyperfine structures and the sequences of energy-levels, also in the particular case of strontium  \cite{Safronova:2010a, *Mani:2010}.
Typical precisions for these calculations can reach the THz for the energy-levels and some tens of MHz for the isotope shifts \cite{Berengut:2003}, therefore experimental data are precious to put constraints on it.
 
From an experimental point of view, the first quantitative data about isotope-shifs of dipole-allowed transitions in Sr$^+$ were obtained by hollow-cathode spectroscopy \cite{Hughes:1957, *Heilig:1961} that reached a precision of 8~MHz using a Doppler-free technique \cite{Lorenzen:1982}.
The development of the method of fast-beam laser spectroscopy in a particle accelerator \cite{Nortershauser:2010} opened the possibility to study a larger number of isotopes (including the unstable ones)  and increased the precision to better than 5~MHz in the most favorable cases.
In the particular case of Sr$^+$, the data made available by this technique concern the $5s\,\, ^2S_{1/2}\to 5p\,\, ^2P_{3/2}$ transition ($\nu=735$~THz, $\lambda=408$~nm) for all the isotopes between $^{78}$Sr and $^{100}$Sr \cite{Buchinger:1990}, as well as some preliminary data for the $5s\,\,^2S_{1/2}\to 5p\,\,^2P_{1/2}$ transition ($\nu=711$~THz, $\lambda=422$~nm) for $^{84}$Sr$^+$, $^{86}$Sr$^+$, and $^{87}$Sr$^+$ \cite{Borghs:1983} (all the shifts are measured with respect to the most abundant isotope $^{88}$Sr$^+$).
This technique also gives reliable information about hyperfine splitting of the odd isotopes with a precision still of the order of 5~MHz.
Fast-beam laser spectroscopy is based on the excitation of a strong dipole-allowed transition from the fundamental state.
This circumstance allows for a direct detection of the fluorescence or for the accumulation in a metastable state that can be state-selectively neutralized or ionized  \cite{Nortershauser:2010}.
While very precious for nuclear-physics, the technique is not well-adapted to the study of dipole-forbidden transitions or transitions from an excited state.
Another possibility to obtain such spectroscopic information is offered by laser-cooled trapped-ion experiments that can provide data with a precision only limited by signal-to-noise issues, the Doppler broadening being negligible.
Some alkali-earth ions \cite{Zhao:1995, *Alt:1997} and  other species \cite{Imajo:1996, *Wang:2007} have been studied by this technique, providing also useful data for quantum-information experiments based on trapped ions.
The use of trapped ions as frequency standards in the optical domain (in particular the $^{88}$Sr$^+$ ion \cite{Margolis:2004, *Barwood:2004, *Madej:2004}) triggered several measurements of the absolute frequency of the dipole-forbidden ``clock'' $5s\,\,^2S_{1/2}\to 4d\,\,^2D_{5/2}$ Sr$^+$ transition ($\nu=446$~THz, $\lambda=674$~nm) together with its isotope shifts $\nu_{88}-\nu_{87}$ down to 40~kHz precision \cite{Barwood:2003}.
A recent paper reports on the  $\nu_{88}-\nu_{86}$ isotope shift in this same clock transition with 4~kHz precision  \cite{Lybarger:2011} .

In the present paper we report the measurements of the isotope shifts for both natural even-isotopes $^{84}$Sr$^+$ and $^{86}$Sr$^+$ with respect to  $^{88}$Sr$^+$.
We address the case of the dipole-allowed transitions $5s\,\,^2S_{1/2}\to 5p\,\,^2P_{1/2}$ ($\nu=711$~THz) and $4d\,\,^2D_{3/2}\to 5p\,\,^2P_{1/2}$ ($\nu=275$~THz, $\lambda=1092$~nm) that we study in laser-cooled trapped-ion samples consisting of two-component Coulomb crystals \cite{Hornekaer:2001}. 
This technique allows us to probe the two isotopes in the same sample during the same frequency-scan, reducing the requirements for long-term laser stabilization needed in the case of sequential experiments \cite{Lybarger:2011}.
The isotope shifts are extracted from the spectra by fitting the experimental data with the solution of the optical Bloch equations (OBE) describing a $\Lambda$ three-level atom in interaction with two laser beams \cite{Stalgies:1998}.
The final precision, estimated by the dispersion of the data, is between 2 and 4~MHz, depending on the isotope and on the transition (in the case of $^{84}$Sr$^+$ two distinct measurement are necessary).
This precision is mainly limited by the laser stabilization technique, 
As mentioned above, such information can inform theoretical models of isotope shifts that are, at the moment, available with larger uncertainties \cite{Berengut:2003, Lybarger:2011}.
This mapping of isotope shifts in the case of Sr$^+$ is useful for quantum information experiments based on this species \cite{Wang:2010a}, especially in the case of multi-isotope sympathetically-cooled ion strings \cite{Home:2009}.

The paper is organized as follows.
In section \ref{sec:exp} we present the experimental setup and we give some details concerning the ion trapping, the frequency-stabilization of the lasers and the isotope enrichment procedure.
Then in section  \ref{sec:theo} we briefly present the theoretical model used to fit the experimental data.
We present the results in section \ref{sec:res} and we compare them to the literature.
Finally, in section \ref{sec:discuss} we discuss the limitations of the present measurements and some possible improvements.

\section{Experimental methods}
\label{sec:exp}
 \subsection{Trapping, cooling, and laser-locking}
Sr$^+$ ions are loaded in a linear Paul trap from an oven containing a strontium dendrite (Aldrich, 99.9\% pure).
Neutral atoms are ionized by driving a two-photon transition towards a self-ionizing level \cite{Removille:2009, *Kirilov:2009}.
The photo-ionizing laser pulses are issued from a frequency doubled Ti:Sa oscillator (Tsunami, Spectra-Physics) with a central frequency of 695~THz ($\lambda=431$~nm) and a pulse duration of $\simeq 100$~fs.
The spectral width associated with these ultrafast pulses ($\simeq 10$~THz) makes this loading process insensitive to both Doppler effect and isotope shift, providing samples with a composition that respects the natural abundances.
The abundances of the four natural isotopes of strontium ($^{84}$Sr, $^{86}$Sr, $^{87}$Sr, $^{88}$Sr) are, respectively 0.56\%, 9.86\%, 7\% and 82.58\%.
The ion trap geometry has been described previously \cite{Removille:2010}.
The trap is driven with a RF voltage amplitude $V_{RF}\sim 750$~V at a frequency of 7.6~MHz and typically displays a radial frequency of 200~kHz for Sr$^+$.
The ``end cap'' electrodes (arranged between the rods and separated by 26~mm) are independently brought to static
voltages $V_{ec}$ of the order of 100~V.
The stray electric-fields can be compensated by adding a different dc voltage on each trap rod, following a technique similar to that of reference~\onlinecite{Herskind:2009}.

\begin{figure}[h]
  \centerline{\includegraphics[width=.75\columnwidth]{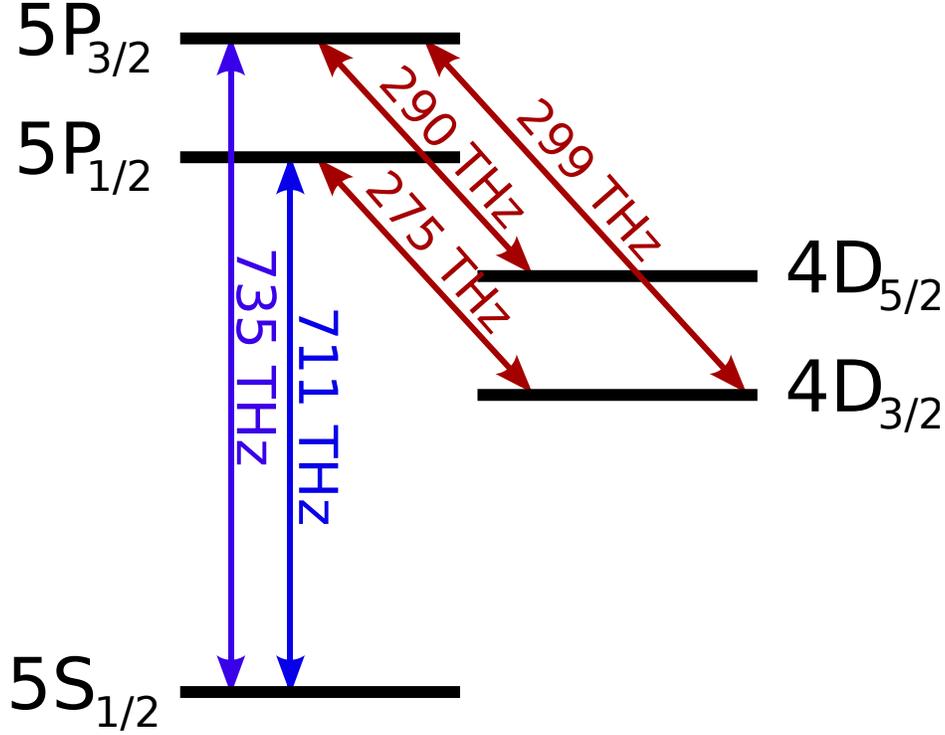}}
  \caption{(Color online) Low energy levels scheme for Sr$^+$. In the present work we address the 711~THz violet transition (cooling) and the 275~THz infrared transition (``repumping'').}
  \label{fig:levels} 
\end{figure}

Trapped Sr$^+$ ions (even isotopes) are Doppler cooled using the 711~THz  $5s\,\,^2S_{1/2}\to 5p\,\,^2P_{1/2}$ optical transition (natural linewidth $\Gamma_b/2\pi=21.5$~MHz, see figure~\ref{fig:levels}).
This transition is driven using laser light generated by a commercial single-mode Ti:Sa CW laser (Coherent MBR~110) frequency-doubled in a single-pass geometry using a periodically-poled KTP crystal (PPKTP, Raicol Crystals). 
Up to 20~mW are available after the doubling process and the coupling into a single-mode polarization-maintaining optical fibre.
The laser frequency is locked to an atomic reference, taking advantage of the near-coincidence ($\nu_{Sr^+}- \nu_{Rb}= 440$~MHz) between the $^{88}$Sr$^+$  $5s\,\,^2S_{1/2}\to 5p\,\,^2P_{1/2}$ and the $^{85}$Rb $5s\,\,^2S_{1/2}(F=2)\to 6p\,\,^2P_{1/2}(F^{\prime}=3)$ transitions \cite{Madej:1998, *Sinclair:2001}.
The 710 962 401 328(40)~kHz absolute frequency of this $^{85}$Rb transition has been recently measured by the frequency-comb technique \cite{Shiner:2007}.

The electronic signal for laser-locking is obtained by a saturated-absorption setup, based on a natural rubidium cell heated to $120$°~C.
The probe-beam signal is demodulated at the first harmonic of the $\sim 90$~kHz frequency-modulation produced by the laser-driver electronics to lock the laser to the intra-cavity thin etalon.
The residual FWHM of the saturated-absorption lines is on the order of 16~MHz, dominated by the residual (unlocked) laser linewidth and modulation (the natural line-width being $\sim 1.3$~MHz \cite{Marek:1980}).
The overall stability of the lock has been evaluated by monitoring the beat note between the Ti:Sa laser and an extended-cavity GaN diode laser (TOPTICA DL100) stabilized with the same technique on the same atomic transition in a different Rb cell and then shifted by an acousto-optic modulator.
The long-term (10 minutes) full width of the beat-note acquired on a spectrum analyzer (2.5~MHz at -3 dB) gives us an upper limit to the performances of the locking method.
Note that all the systematic error contributions evaluated in a similar setup \cite{Shiner:2007} are largely smaller than these fluctuations.

A commercial fiber-laser (Koheras Adjustik Y10) drives the $4d\,\,^2D_{3/2}\to 5p\,\,^2P_{1/2}$ 275~THz ``repumping'' transition (see figure~\ref{fig:levels}) to avoid the accumulation of the ions into the metastable $4d\,\,^2D_{3/2}$ state during the cooling process.
This laser has a nominal linewidth of 70~kHz and it is stabilized against long term drifts by a transfer-lock technique using a scanning ring cavity referenced to a stabilized 711~THz laser-diode \cite{Burke:2005, Seymour-Smith:2010}.
This feedback loop is, in our case, relatively slow (bandwidth $\simeq 3$~Hz).
Nevertheless, the computer-controlled lock allows us to scan the 275~THz fiber laser over many free-spectral ranges of the ring cavity [229.2(4)~MHz].
We estimated the statistical uncertainty on the frequency of this laser by acquiring during several tens of hours the error signal.
The histograms show a gaussian behavior with FWHM of 1.2~MHz; this quantity gives us an upper limit to the slow drift error on the ``two-photon detuning'' $\delta_b-\delta_r$ (see figure \ref{fig:levelstheo} below).
Ring-cavity piezo-actuator nonlinearities are corrected for dynamically (i.e. for each acquisition of a ring-cavity scan) in order to obtain a linear and calibrated frequency scan. After correction we estimate at 500~kHz the upper limit of the residual systematic shift affecting two frequencies separated by a free spectral range of the cavity.

A magnetic field of  the order of $0.3\times 10^{-4}$~T defines a quantization-axis along the trap axis, orthogonal with respect to the linear polarization of the repumping and cooling laser beams.
This configuration prevents the ions from being optically-pumped into a metastable dark state by the repumping laser alone \cite {Berkeland:2002a}.
However, it does not prevent the existence of ``two-colour'' dark states that causes a dip in the infrared spectra for $\delta_r=\delta_b$ \cite{Janik:1985, *Siemers:1992} (see figure \ref{fig:levelstheo} below) .
As explained in the following, we take advantage of this circumstance to measure violet and infrared isotope shifts in a single frequency scan.
The presence of a longitudinal magnetic field induces a Zeeman splitting of the $5s\,\,^2S_{1/2}$, $4d\,\,^2D_{3/2}$ and $5p\,\,^2P_{1/2}$ levels.
As explained below, these splittings induce a broadening of the (unresolved) dark resonance of the order of 2~MHz.

\subsection{Isotope enrichment and addressing}
After the loading of the trap, the composition of the Doppler-cooled Coulomb crystal can be modified by taking advantage of both the radial segregation induced by the mass differences 
and the axial segregation induced by the radiation pressure of the axial cooling beams.
For example, as described in detail elsewhere \cite{Dubost:2012, *Dubost:2013}, by lowering the potential of the endcap corresponding to the input side of the cooling beams it is possible to selectively remove the uncooled species (dark ions) that are not pushed towards the other side of the trap. 
Such a strategy requires the laser addressing of at least two different isotopes [for both cooling (711~THz) and repumping (275~THz) transitions].
By this technique we routinely obtain almost pure two-isotope Coulomb crystals consisting of up to  $10^4$ $^{88}$Sr$^+$ +  $10^4$ $^{86}$Sr$^+$ ions or   $10^3$ $^{86}$Sr$^+$  + $10^3$ $^{84}$Sr$^+$ ions.

\begin{figure}[h]
  \centerline{\includegraphics[width=\columnwidth]{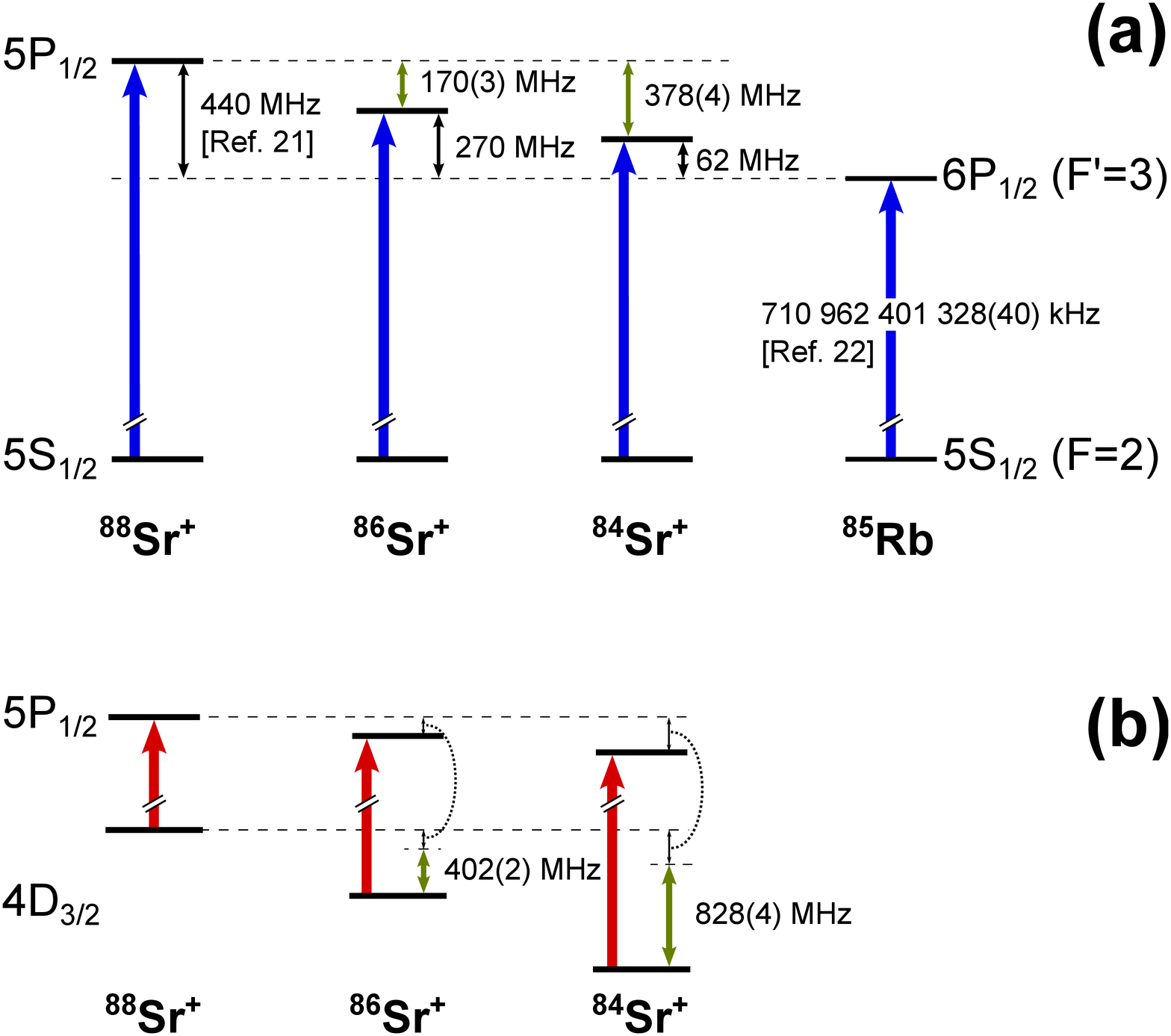}} 
 \caption{(Color online) Schematic drawing of the isotope shifts of the even Sr$^+$ isotopes studied in this paper (color online). (a) $5s\,\,^2S_{1/2}\to 5p\,\,^2P_{1/2}$ violet transition (711~THz , blue arrows): green arrows represent the measured isotope shifts while black arrows indicate the frequency shift with respect to the $^{85}$Rb $5s\,\,^2S_{1/2}(F=2)\to 6p\,\,^2P_{1/2}(F^{\prime}=3)$ reference transition. Note that we referenced the energies to the ground states. (b) $4d\,\,^2D_{3/2}\to 5p\,\,^2P_{1/2}$ infrared transition (275~THZ, red arrows): green thick arrows represent the measured isotope shifts with respect to $^{88}$Sr$^+$ ``repumping'' transition. As in this diagram both levels are isotope-shifted, we show also with thin black arrows the isotope shift that affects the 711~THz transition. The errors given are an estimate of the statistical variance in our measurements.}
  \label{fig:shifts}
\end{figure}

The 711~THz beam (frequency-locked to the $^{85}$Rb transition) is split into two separate beams in order to address simultaneously two isotopes (either $^{88}$Sr$^+$+$^{86}$Sr$^+$ or $^{86}$Sr$^+$+$^{84}$Sr$^+$).
Each of these beams is then frequency up-shifted by a double-pass acousto-optical modulator (AOM).
When addressing a $^{88}$Sr$^+$+$^{86}$Sr$^+$ Coulomb crystal, the first AOM is driven by a RF source with a frequency of 220~MHz, the second one with a frequency of 135~MHz.
The 275~THz laser is also split into two beams that are frequency up and down-shifted by $\simeq 200$~MHz respectively. 
As shown in figure~\ref{fig:shifts}, these frequency shifts allow us to address transitions of both $^{88}$Sr$^+$ and $^{86}$Sr$^+$ and to set detunings $\delta_b$ and $\delta_r$ (figure~\ref{fig:levelstheo}) within a band of, respectively, $\pm 50$~MHz and $\pm 60$~MHz.

While addressing the pair $^{86}$Sr$^+$+$^{84}$Sr$^+$, the 711~THz laser is locked instead to the  $^{85}$Rb $5s\,\,^2S_{1/2}(F=2)\to 6p\,\,^2P_{1/2}(F^{\prime}=2)$ transition, red-detuned by 117 366(72)~kHz with respect to the $F=2\to F^\prime=3$ transition \cite{Shiner:2007}.
In this way, driving the first violet AOM with $\simeq 180$~MHz and the second with $\simeq 85$~MHz allows us to address the cooling transitions (see figure~\ref{fig:shifts}); however a shorter detuning range is available.
The repumping beams can be simply tuned to the new isotope pair by red-shifting the master 275~THz laser by $\sim 400$~MHz and slightly 
adjusting the AOM frequencies.
The final precision of the isotope shift measurements relies on the precision of the AOM driving frequency, based on voltage-controlled oscillators (VCO).
In order to take into account the possible drifts of the RF drivers, we recorded the frequencies of all the AOM drivers (measured with a spectrum analyzer) during each acquisition of a spectrum.
The precision of these frequency shifts is then limited by the bandwidth of the measurement on the spectrum analyzer ($\sim 50$~kHz), the absolute calibration error of the instrument being negligible at that level.

 \subsection{Fluorescence spectra\label{sec:spectra}}
After loading the trap and applying the isotope-selection procedure, we acquire fluorescence spectra. 
The fluorescence signal is obtained by imaging the whole ion crystal on a CCD camera (Prosilica GC1600) with a $f=58$~mm standard camera lens. 
Typical integration times are of the order of 1 second; the images are corrected for background that is acquired in the absence of ions at the beginning and at the end of a spectrum.
Spectra are acquired in an interlaced way following this sequence:
\begin{itemize}
\item the detuning of the master 275~THz laser is changed
\item both isotopes are cooled during 500~ms
\item the lasers addressing the first isotope are switched off while the second isotope cycles in the cooling transition; one fluorescence image is taken
\item the lasers addressing the second isotope are switched off while the first isotope cycles in the cooling transition; one fluorescence image is taken
\end{itemize}
This procedure reduces the susceptibility of the isotope shift measurement to possible long-term drifts of the locked lasers. 
Moreover, sympathetic cooling makes the ion loss-rate indistinguishable from that measured in a continuous cooling regime (lifetime $\simeq$~3~hours for the whole crystal).

In the experiments we obtained in a single spectral determination both the isotope shifts of the $5s\,\,^2S_{1/2}\to 5p\,\,^2P_{1/2}$ Sr$^+$ and $4d\,\,^2D_{3/2}\to 5p\,\,^2P_{1/2}$ Sr$^+$ transitions.
Fluorescence spectra were recorded by setting the cooling laser frequency at a moderate red detuning  $\delta_b\simeq -30$~MHz and by frequency-scanning the 275~THz laser beam moving the locking point of the transfer lock.
In this way the repumpers of both isotopes are scanned simultaneously with a mutual detuning determined by the frequencies of the two AOM drivers that are kept fixed during the experiment.
Experiments were repeated several times for different repumper intensities and $\delta_b$ in order to verify the reproducibility and to obtain an estimate of statistical errors.
Moreover, we also fixed a detuning on AOMs far from the isotope shift matching condition: in this way the spectra of the two isotopes appear shifted by a non-negligible amount that should allow for the detection of crosstalk in the fluorescence recordings that, however, has not been observed.
As mentioned before, the spectra obtained in this way display a dip for $\delta_r=\delta_b$ \cite{Janik:1985, *Siemers:1992}.
This spectral feature gives us the opportunity to measure in a single spectrum the position of both transitions.
This information is extracted from the experimental spectra by fitting the data with a theoretical model, as described in the next section.

\section{Theoretical model for the fluorescence spectra}
\label{sec:theo} 

Optical Bloch equations (OBE) within a three level model describe the behavior of a $\Lambda$ atomic system driven by two lasers occurring, for example, in laser-cooled alkali-earth ions \cite{Stalgies:1998}.
In brief, after applying the rotating-wave approximation and taking into account the branching ratios and decay-rates from the literature ($\Gamma_b/2\pi=21.5$~MHz, branching ratio=13.4 \cite{Gallagher:1967}), the steady state solution of the resulting equation can be analytically obtained by matrix-inversion. 
As the transients are short compared to the averaging time, this procedure allows us to fit the experimental data (fluorescence vs $\delta_r$) with the steady state solution of the OBE, having the following free parameters: $\Omega_b$ (resp. $\Omega_r$), the Rabi-frequency at resonance of the cooling (resp. repumping) laser;  $\delta_b$, the detuning of the cooling laser; $\gamma_{DS}$, the decoherence rate between $D$ and $S$ levels; $\omega_{0r}$, the central angular frequency of repumping transition. The decoherence rate $\gamma_{PS}$ ($\gamma_{PD}$) between $P$ and $S$ ($P$ and $D$) levels is fixed to $\Gamma_b /2$.   

\begin{figure}[h]
  \centerline{\includegraphics[width=.75\columnwidth]{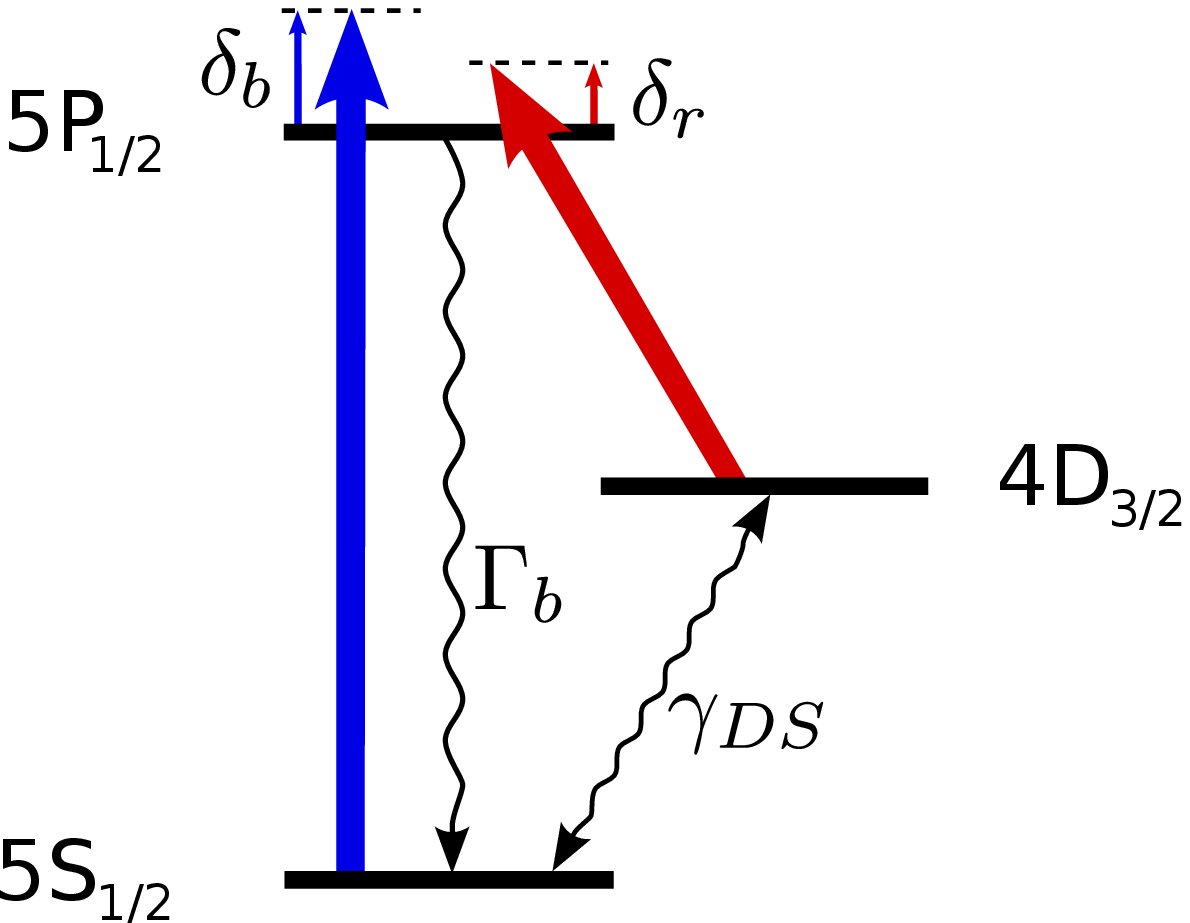}}
  \caption{(Color online) Schematic drawing of the atomic levels of Sr$^+$ ($\Lambda$ configuration). One laser (``cooling'') addresses the $S\to P$ transition with a detuning  $\delta_b$ and a Rabi frequency $\Omega_b$ while another (``repumping'') laser drives the $D\to P$ transition (detuning $\delta_r$ and Rabi frequency $\Omega_r$). The decoherence rate between $D$ and $S$ levels is given by $\gamma_{DS}$.}
  \label{fig:levelstheo}
\end{figure}

\begin{figure}[h]
\centerline{\includegraphics[width=.99\columnwidth]{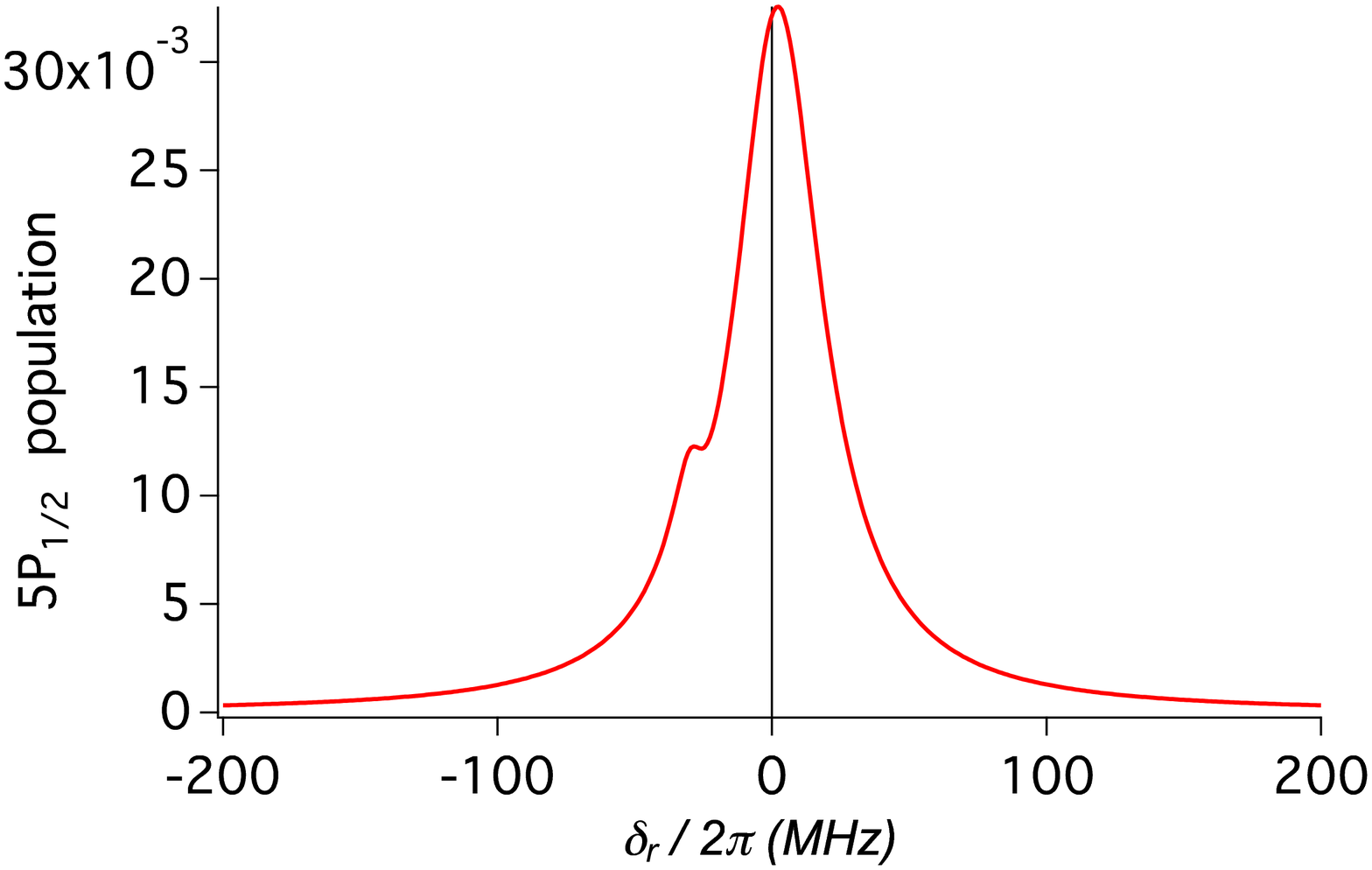}}
 \caption{(Color online) Theoretical fluorescence spectrum as a function of $\delta_r$ (Optical Bloch Equations). The parameters of the plot are: $\Gamma_b/2\pi=21.5$~MHz; branching ratio=13.4 \cite{Gallagher:1967}; $\delta_b/2\pi=-28$~MHz; $\Omega_b/2\pi=12.9$~MHz; $\Omega_r/2\pi=2$~MHz; $\gamma_{DS}/2\pi=5.7$~MHz. These parameters correspond to the fit of the rightmost spectrum in figure~\ref{fig:redspectra}}
 \label{fig:theospectrum} 
\end{figure}

The three-level approximation neglects multiple Zeeman sublevels but allows us to take into account the light-shifts and to reproduce the shape of the experimental spectra.
The output from the OBE (figure~\ref{fig:theospectrum}) has been also exploited in order to optimize the experimental parameters: when decreasing $\Omega$, the spectra become narrower and should allow for a more precise determination of the $\delta_r=0$ and $\delta_r=\delta_b$ conditions but, in the experiment, signal-to-noise ratio decreases down to very poor levels.

Let us now discuss the role of the parameter $\gamma_{DS}$ that affects both the width and the contrast of the dip in the spectra.
Taking into account the lifetime of the metastable $4 ^2D_{3/2}$ state \cite{Mannervik:1999} we expect $\gamma_{DS}/2\pi=0.18$~Hz.
The fit procedure adjusts $\gamma_{DS}$ in order to take into account the experimental shape of the dip that is, however, affected by several additional factors: the relative phase noise of the lasers, the presence of a magnetic field that mixes coupled and uncoupled states \cite{Berkeland:2002a}, and the non-zero temperature of the sample.
We also note that the axial magnetic field should split the dip, the system being no longer correctly described by a three-level model \cite{Janik:1985, *Siemers:1992}.
In our case, however, this splitting is negligible with respect to the observed linewidth, in the 5~MHz range.

We emphasize that this width, by itself, does not put a fundamental limit to the accuracy of the measured isotope shifts, as long as systematic shifts are negligible and signal to noise ratio high enough.
Systematic shifts of the parameters $\delta_b$ and $\omega_{0r}$ arise through Doppler broadening induced by residual axial ion motion, and we checked this effect by retrieving these parameters from broadened spectra obtained by convolution of the theoretical line-shape with a Maxwell-Boltzmann velocity distribution (temperature $T$).
We estimate this systematic shift to be $2\pi\times -4.7$~MHz ($\omega_{0r}$ ) and  $2\pi\times +19$~MHz ($\delta_b$) at a $T=700$~mK, corresponding to liquid-gas transition of the ion sample \cite{Dubin:1999} that clearly overestimates the actual temperature of the ions ensembles in the experiment.
However the isotope shift is obtained by taking the difference between the parameters $\delta_b$ and $\omega_{0r}$ of two simultaneously acquired spectra.
In such a procedure, the "common mode" shifts induced by a finite temperature cancel out and the efficient sympathetic coupling is able to hold the differential temperature $\Delta T$ between species down to the 10~mK regime \cite{Bowe:1999}. 
We calculate that differential temperatures should rise to 100~mK in order to induce systematic shifts larger than the statistical errors of the experiment.

\section{Results}
\label{sec:res}

\begin{figure}[h]
  \centerline{\includegraphics[width=.75\columnwidth]{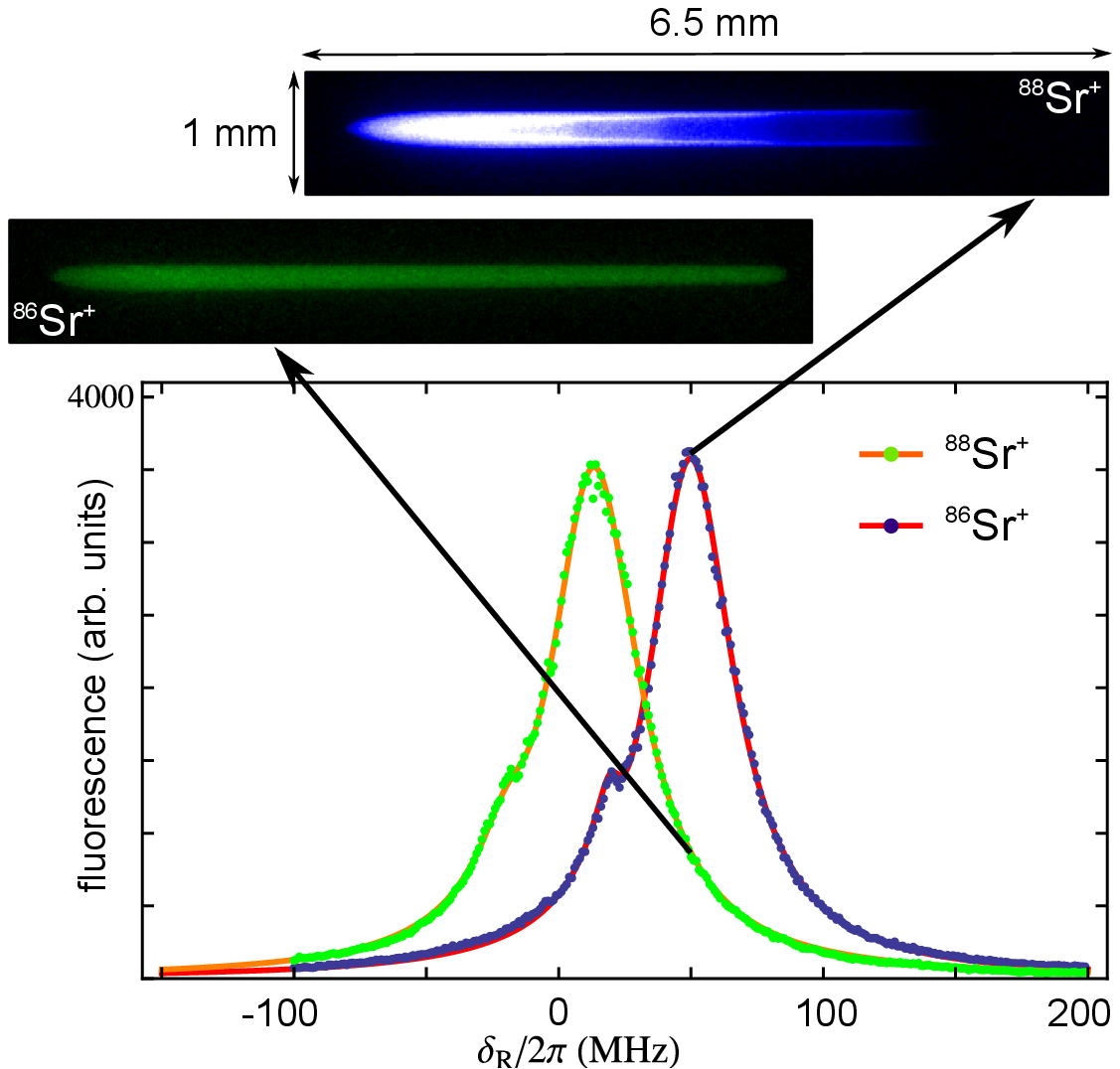}}
  \caption{(Color online) Twin spectra obtained with a Coulomb crystal composed of equivalent proportions of $^{88}$Sr$^+$ and $^{86}$Sr$^+$ ions. The repumping laser detunings $\delta_r$ of the two isotopes are scanned simultaneously but there is a constant frequency shift between the two beams imposed by fixed-frequency acousto-optic modulators ( $\sim 40$~MHz in this example). For each detuning of the master 275~THz laser two images are acquired. As an example we show two of such images coded in false colors corresponding to the $^{88}$Sr$^+$ (blue) and $^{86}$Sr$^+$ (green) isotopes. The effect of radial segregation in the trap pushes the $^{88}$Sr$^+$ in the outer shells, the core of the crystal being occupied by $^{86}$Sr$^+$. The total fluorescence signal plotted in the graph is obtained by integrating these images in the region of interest. The continuous lines are the best fit of the spectra by an analytical model based on a three-level atom (optical Bloch equations). The fitting parameters for the $^{88}$Sr$^+$ spectrum are listed in the caption of figure~\ref{fig:theospectrum}; the fitting parameters of the $^{86}$Sr$^+$ spectrum are: $\Gamma_b/2\pi=21.5$~MHz; $\delta_b/2\pi=-25.9$~MHz; $\Omega_b/2\pi=16.1$~MHz; $\Omega_r/2\pi=2.7$~MHz; $\gamma_{DS}/2\pi=8.6$~MHz.}
  \label{fig:redspectra}
\end{figure}
As explained above, the isotope shift determination is based on the acquisition of simultaneous infrared spectra on two-isotope crystals, scanning the $4d\,\,^2D_{3/2}\to 5p\,\,^2P_{1/2}$ Sr$^+$ transition.
A typical example of twin spectra obtained with this technique is shown in figure~\ref{fig:redspectra} for the pair $^{88}$Sr$^+$ and $^{86}$Sr$^+$.
The continuous line is the best fit with the OBE using the six free parameters: $\Omega_b$, $\Omega_r$, $\gamma_{DS}$, $\delta_b$, $\omega_{0r}$ and a scale factor that takes into account the photon collection efficiency.
The model is able to reproduce all the main features of the experimental spectra.
The $\Omega$ parameters obtained from fit agree within a factor of 3 with intensity and beam-size measurements.
The $\delta_b$ and $\omega_{0r}$ parameters can be used (knowing the fixed frequency-shift imposed by the AOMs) to retrieve the isotope shifts in both the $5s\,\,^2S_{1/2}\to 5p\,\,^2P_{1/2}$ Sr$^+$ and $4d\:^2D_{3/2}\to 5p\:^2P_{1/2}$ Sr$^+$ transitions.

\begin{table}[h]
\[
\begin{array}{|c|}
\hline
5s\:^2S_{1/2}\to 5p\:^2P_{1/2}\mbox{~transition~}(711\mbox{~THz})\\
\begin{array}{c|c|c|c}
\hline
\mbox{isotope} & \Delta_{88} \mbox{(MHz)} & \begin{array}{c}\mbox{previous}\\\mbox{result}\\\end{array} & \mbox{ref.}\\
\hline  ^{86}\mbox{Sr}^+  & +170(3) & +168(8)& \mbox{\cite{Lorenzen:1982}}\\
\hline  ^{84}\mbox{Sr}^+  & +378(4) & +378(12)& \mbox{\cite{Heilig:1961}}\\
\end{array}\\
\hline
\end{array}
\]
\caption{\label{tab:redblueshifts} Isotope shifts $\Delta_{88}=\nu_{^{88}Sr^+}-\nu_{^{}Sr^+}$  for the two less abundant even isotopes as obtained by frequency-scanning the repumping laser beam. The shifts were retrieved by fitting the experimental data to a three-level based OBE ($\delta_b$ parameters) and by taking into account the fixed AOM shifts imposed to the 711~THz beams. The 88--86 shift comes from a direct measurement while the 88--84 shift is based on two distinct measurements (the 88--86 shift and the 86--84 shift).}
\end{table}

\begin{table}[h]
\[
\begin{array}{|c|}
\hline
4d\:^2D_{3/2}\to 5p\:^2P_{1/2}\mbox{~transition~}(275\mbox{~THz})\\
\hline
\begin{array}{c|c}
\qquad\mbox{isotope}\qquad& \qquad\Delta_{88} \mbox{(MHz)}\qquad \\
\hline  ^{86}\mbox{Sr}^+ & -402(2) \\
\hline  ^{84}\mbox{Sr}^+ & -828(4) \\
\end{array}\\
\hline
\end{array}
\]
\caption{\label{tab:redshifts} Isotope shifts $\Delta_{88}=\nu_{^{88}Sr^+}-\nu_{^{}Sr^+}$  for the two even isotopes as obtained by frequency-scanning the repumping laser beam. The frequency difference for the $4d\,\,^2D_{3/2}\to 5p\,\,^2P_{1/2}$ Sr$^+$ transitions was retrieved by fitting the experimental data to a three-level based OBE ($\omega_{0r}$ parameters) and by taking into account the fixed AOM shifts. The 88--86 shift comes from a direct measurement while the 88--84 shift is based on two distinct measurements.}
\end{table}

As the points in the two spectra are acquired in an interleaved way, the precision of the isotope shift is not affected by the possible slow drift of the infrared laser but only by the fast fluctuation not corrected by the frequency-lock.
It was easy to check the validity of this approach by imposing a non-negligible shift between the twin spectra (change in the frequency of one of the 275~THz AOMs, see figure~\ref{fig:redspectra}) or by varying other parameters (mainly laser intensities).
We analyze here 12 spectra for the pair $^{88}$Sr$^+$ + $^{86}$Sr$^+$ and 7 spectra for the pair $^{86}$Sr$^+$ + $^{84}$Sr$^+$.
The data extracted from the spectra are summarized in tables~\ref{tab:redblueshifts} and \ref{tab:redshifts}, 
with the estimated (statistical and systematic) error.
Isotope shifts are also compared (whenever possible) to the most precise previously available determinations.

\section{Discussion}
\label{sec:discuss}
When compared to existing determinations of isotope shifts, the data that we obtained using cold trapped ions are in excellent agreement with the previous measurements based on very different techniques.
The new determinations of the isotope shifts in dipole-allowed transition from the metastable state will allow for more stringent tests of the theory.
This technique could be easily used to measure the isotope shift of the two other low-lying transitions in Sr$^+$: the $4d\,\,^2D_{5/2}\to 5p\,\,^2P_{3/2}$ at 290~THz and the $5s\,\,^2S_{1/2}\to 5p\,\,^2P_{3/2}$ at 735~THz.
In order to do this, one needs to address these transitions with stabilized lasers and to add a repumping beam on the $4d\,\,^2D_{3/2}\to 5p\,\,^2P_{1/2}$ transition (see figure~\ref{fig:levels}).

The observed statistical noise that affects the isotope shift measurements in the present work is dominated by the residual fluctuations of laser frequency due to poor locking techniques.
Following reference~\onlinecite{Seymour-Smith:2010}, it would be possible to implement a faster (and tighter) transfer lock for the 275~THz laser.
These authors report a stability better than 10~kHz ($t>1s$) that would bring the lock stability of the infrared laser down to negligible levels.  
The locking of the 711~THz laser could also be improved.
The authors of reference~\onlinecite{Shiner:2007} report a 40~kHz absolute precision when locking the laser to the Rb atomic transition.
Improvements in this locking technique should then also allow for an increased precision.
In particular, an offset-lock of the cooling laser to a Rb-referenced laser could also expand the possible detuning-ranges while addressing different isotopes. 
Finally, reducing the number of ions used in the experiment should drastically reduce the systematic effects of residual magnetic-field (easier to compensate in a point-like geometry) and ion motion (micromotion, residual temperature).
Single-ion spectra in an ion-chain containing two isotopes should be therefore the optimal geometry for a more precise determination of the isotope shifts.
Moreover, in this configuration, the use of sympathetic cooling by another ion \cite{Barrett:2003} could drastically reduce the residual heating effect arising in the positive-slope part of the dip of infrared spectra \footnote{D. M. Lucas, private communication}.
This could reduce systematic error (deformation and shift of the spectra) in this kind of experiments.

\section*{Acknowledgements}
We thank M. Apfel and P. Lepert for technical support.
We also thank S. Removille, Q. Glorieux and T. Coudreau for fruitful discussions.
We are grateful to D. M. Lucas for a critical reading of the manuscript and for enlightening discussions. 
This study was partly founded by Région Ile-de-France through a SESAME project.
\bibliography{/Users/luca/Documents/biblio_total}

\end{document}